\documentstyle[b98proc,epsf]{article}

\def\Journal#1#2#3#4{{#1} {\bf #2}, #3 (#4)}

\def\NPB{{\em Nucl. Phys.} B}

\def\PL{{\em Phys. Lett.}}

\def\PREV{\em Phys. Rev.}

\begin{document}

\def\la{\mathrel{\mathpalette\fun <}}
\def\ga{\mathrel{\mathpalette\fun >}}
\def\fun#1#2{\lower3.6pt\vbox{\baselineskip0pt\lineskip.9pt
\ialign{$\mathsurround=0pt#1\hfil##\hfil$\crcr#2\crcr\sim\crcr}}}

\title{
PROTON SPIN CONTENT, AXIAL COUPLING CONSTANTS AND QCD TOPOLOGICAL SUSCEPTIBILITY
}

\author{B. L. IOFFE, A.G. OGANESIAN}
\address{\it Institute of Theorertical and Experimental Physics,\\
Moscow 117218, Russia}

% You may repeat \author and \address if necessary

\maketitle

\vspace{12pt}
% beginning of abstract

\abstracts{
The part of the proton spin $\Sigma$ carried by $u, d, s$ quarks is
calculated in the framework of the QCD sum rules in the external
fields. The operators up to dimension 9 are accounted.  An important
contribution comes from the operator of dimension 3, which in the limit
of massless $u, d, s$ quarks is equal to the derivative
of QCD topological  susceptibility $\chi^{\prime} (0)$. The comparison
with the experimental data on $\Sigma$ gives $\chi^{\prime}(0)= (2.3
\pm 0.6) \times 10^{-3} ~ GeV^2$. The limits on $\Sigma$ and
$\chi^{\prime}(0)$ are found from selfconsistency of the sum rule,
$\Sigma = 0.33\pm 0.20$, $\chi^{\prime}(0) = (2.4\pm 0.5)\cdot 10^{-3}~GeV.$
The values of $g_A = 1.37 \pm 0.10$ and $g^8_A = 0.65 \pm 0.15$ are also
determined.}

%\end{document}
%\newpage

%\large

%\vspace{8mm}

We calculated \cite{OGA} the value of $\Sigma$ -- the part of
nucleon spin carried by three flavours of light quarks
$\Sigma = \Delta u + \Delta d + \Delta s$,
where $\Delta u, \Delta d, \Delta s$ are the parts of nucleon spin
carried by $u,d,s$ quarks. On the basis of the operator product expansion
(OPE) $\Sigma$ is related to the proton matrix element of the flavour
singlet axial current $j^0_{\mu 5}$
%1
\begin{equation}
2 ms_{\mu} \Sigma = \langle p, s \vert j^0_{\mu 5} \vert p,s \rangle,
\end{equation}
where $s_{\mu}$ is the proton spin 4-vector, $m$ is the proton mass.
The polarization operator
%2
\begin{equation}
\Pi(p) = i \int d^4 x e^{ipx} \langle 0 \vert T \{ \eta(x),
\bar{\eta} (0) \} \vert 0 \rangle
\end{equation}
was considered, where  $\eta(x)$
is the current with proton quantum numbers \cite{IOF}.
It is assumed that the term $\Delta L = j^0_{\mu 5}A_{\mu}$
where $A_{\mu}$ is a constant singlet axial field, is added to QCD Lagrangian.
In the weak axial field approximation $\Pi(p)$ has the form
%3
\begin{equation}
\Pi(p) = \Pi^{(0)} (p) + \Pi^{(1)}_{\mu} (p) A_{\mu}.
\end{equation}
$\Pi^{(1)}_{\mu}(p)$ is calculated in QCD by
OPE at $p^2 < 0, \vert
p^2 \vert \gg R^{-2}_c$, where $R_c$ is the confinement radius. On the
other hand,  using dispersion relation, $\Pi^{(1)}_{\mu} (p)$ is
represented by the contribution of the physical states, the lowest of
which is the proton state. The contribution of excited states is
approximated as a continuum and suppressed by the Borel transformation. The
desired answer is obtained by equalling of these two representations.

An essential ingredient of the method is the appearance of induced by
the external field vacuum expectation values (v.e.v). The most
important of them in the problem at hand is
%4
\begin{equation}
\langle 0 \vert j^0_{\mu 5} \vert 0 \rangle_A \equiv 3 f^2_0 A_{\mu}
\end{equation}
of dimension 3. The constant $f^2_0$ is related to QCD topological
susceptibility. We can write
%8
$$
\langle 0 \vert j^0_{\mu 5} \vert 0 \rangle_A = lim_{q \to 0} ~ i \int~
d^4 xe^{iqx} \langle 0 \vert T \{ j^0_{\nu 5} (x), j^0_{\mu 5} (0) \}
\vert 0 \rangle A_{\nu} \equiv
$$
\begin{equation}
\equiv lim_{q \to 0} P_{\mu \nu} (q) A_{\nu}
\end{equation}
The general structure of $P_{\mu \nu} (q)$ is
%9
\begin{equation}
P_{\mu \nu} (q) = -P_L(q^2) \delta_{\mu \nu} + P_T(q^2) (-\delta_{\mu
\nu} q^2 + q_{\mu} q_{\nu})
\end{equation}
Because of anomaly there are no massless states in the spectrum of the
singlet polarization operator $P_{\mu \nu}$ even for massless quarks.
$P_{T,L}(q^2)$ also have no kinematical singularities at $q^2 = 0$ .
Therefore, the nonvanishing value $P_{\mu \nu} (0)$ comes entirely from
$P_L(q^2)$. Multiplying $P_{\mu \nu} (q)$ by $q_{\mu} q_{\nu}$, in the
limit of massless $u, d, s$ quarks we get
%7
$$
q_{\mu} q_{\nu} P_{\mu \nu} (q) = -P_L (q^2) q^2 = N^2_f
(\alpha_s / 4 \pi)^2 i \int~ d^4 xe^{iqx} \times
$$
\begin{equation}
\times \langle 0 \vert T {G^n_{\mu \nu} (x) \tilde{G}^n_{\mu \nu} (x),
G^m_{\lambda \sigma} (0) \tilde{G}^m _{\lambda \sigma} (0)} \vert 0 \rangle,
\end{equation}
where $G^n_{\mu \nu}$ is the gluonic field strength, $\tilde{G}_{\mu
\nu} = (1/2) \varepsilon_{\mu \nu \lambda \sigma} G_{\lambda \sigma}$.
At  $q^2 \to 0$, we have
%8
\begin{equation}
f^2_0 = -(1/3) P_L(0) = \frac{4}{3} N^2_f \chi^{\prime} (0),
\end{equation}
where $\chi(q^2)$ is the topological susceptibility
%9
\begin{equation}
\chi(q^2) = i~ \int d^4 x e^{iqx} \langle 0 \vert T {Q_5 (x),
Q_5 (0)} \vert 0 \rangle
\end{equation}
%10
\begin{equation}
Q_5(x) = (\alpha_s / 8 \pi)~ G^n_{\mu \nu} (x) \tilde{G}^n _{\mu
\nu} (0),
\end{equation}
$\chi(0) = 0$ if there is
at least one massless quark \cite{CRE}.

In ref. \cite{KHO} the sum rule, expressing $\Sigma$ in terms of
$f^2_0$ (4) or $\chi^{\prime}(0)$ was found.  The OPE up to
dimension $d=7$ was performed.

However, the
accuracy of the calculation was not good enough for reliable
calculation of $\Sigma$ in terms of $f^2_0$: the necessary requirement
of the method -- the weak dependence of the result on the Borel
parameter was not well satisfied.

In this paper we improve the accuracy of the calculation by going to
higher order terms in OPE up to dimension 9 operators. Under the
assumption of factorization -- the saturation of the product of
four-quark operators by the contribution of an intermediate vacuum
state -- the dimension 8 v.e.v.'s are accounted (times $A_{\mu}$):
%11
\begin{equation}
-q\langle 0 \mid \bar{q} \sigma_{\alpha\beta} (1/2)\lambda^n
G^n_{\alpha\beta} q \cdot \bar{q}q \mid 0\rangle = m^2_0 \langle 0\mid
\bar{q}q \mid 0\rangle^2
\end{equation}
where $m^2_0=0.8\pm 0.2~GeV^2$ was determined in \cite{BEL}
The sum rule for $\Sigma$ is given by
%18
$$
\Sigma + C_0 M^2 = -1 + \frac{8}{9 \tilde{\lambda}^2_N} e^{m^2/M^2}
\left \{a^2 L^{4/9} + 
 + 6\pi^2 f^2_0 M^4 E_1 \Biggl (\frac{W^2}{M^2} \Biggr ) L^{-4/9} +
\right.$$
\begin{equation}
\left. + 14 \pi^2 h_0 M^2 E_0 \Biggl (\frac {W^2}{M^2} \Biggr ) L^{-8/9} -
\frac{1}{4}~ \frac{a^2 m^2_0}{M^2} - \frac{1}{9} \pi \alpha_s f^2_0~
\frac{a^2}{M^2} \right \}
\end{equation}

Here $M^2$ is the Borel parameter, $\tilde{\lambda}_N$ is defined as
$\tilde{\lambda}^2_N = 32 \pi^4 \lambda^2_N = 2.1 ~ GeV^6$,
$\langle 0 \vert \eta \vert p \rangle = \lambda_N v_p,$
where $v_p$ is proton spinor, $W^2$ is the continuum threshold, $W^2 =
2.5 ~GeV^2$,  $h_0=3\cdot 10^{-4}~GeV^4$ \cite{KHO},
%13
\begin{equation}
a = -(2 \pi)^2 \langle 0 \vert \bar{q} q \vert 0 \rangle = 0.55 ~ GeV^3
\end{equation}
$$
E_0(x) = 1 - e^{-x} , ~~~ E_1(x) = 1 - (1 + x)e^{-x}
$$
$L = ln (M/\Lambda)/ ln (\mu/\Lambda),~~~
\Lambda = \Lambda_{QCD} = 200 ~ MeV$ and the normalization point  $\mu$
was chosen $\mu = 1 ~ GeV$. When deriving (12) the sum rule for the
nucleon mass was exploited what results in  appearance of the first
term, -1, in the right hand side (rhs) of (12). This term absorbs the
contributions of the bare loop, gluonic condensate as well as $\alpha_s$
corrections to them and essential part of terms, proportional to $a^2$ and
$m^2_0 a^2$.  The values of the parameters, $a, \tilde{\lambda}^2_N, W^2$
taken above were chosen by the best fit of the sum rules for the nucleon
mass (see \cite{SMI}, Appendix B) performed at $\Lambda = 200 ~ MeV$.
The unknown constant $C_0$ in the
left-hand side  of (12) corresponds to the contribution of inelastic
transitions $p \to N^* \to interaction~ with A_{\mu} \to p$ (and in inverse
order).
The sum rule (12) as well as the sum
rule for the nucleon mass is reliable in the interval of the Borel parameter
$M^2$
$0.85 < M^2 < 1.4 ~ GeV^2$.The $M^2$-dependence of the
rhs of (12) at $f^2_0 = 3 \times 10^{-2}~ GeV^2$ is plotted in Fig.1a. The
complicated expression in rhs of (12) is indeed an almost linear function of
$M^2$ in the given interval!
The best values of $\Sigma = \Sigma^{fit}$ and $C_0 = C^{fit}_0$ are found from the $\chi^2$
fitting procedure
%14
\begin{equation}
\chi^2 = \frac{1}{n} \sum \limits ^{n}_{i=1}~ [\Sigma^{fit} -
C^{fit}_0 M^2_i - R(M^2_i)]^2 = min,
\end{equation}
where $R(M^2)$  is the rhs of (12).

The values of $\Sigma$ as a function of $f^2_0$  are
plotted in Fig.1b together with $\sqrt{\chi^2}$.
In our approach the gluonic
contribution cannot be separated and is included in $\Sigma$. The
experimental value of $\Sigma$ can be estimated \cite{{ADA},{ABE}}
as $\Sigma = 0.3 \pm 0.1$. Then from Fig.1b we have $f^2_0 =
(2.8 \pm 0.7) \times 10^{-2} ~ GeV^2$ and $\chi^{\prime}(0) = (2.3 \pm 0.6)
 \times 10^{-3}~  GeV^2$ . The error in $f^2_0$ and $\chi^{\prime}$ besides
the experimentall error includes the uncertainty in the sum rule estimated
as equal to the contribution of the last term in OPE (two last terms in
Eq.12)
and a possible role of NLO $\alpha_s$ corrections.

From $\chi^2$ fit, i.e. from the requirement of the selfconsistancy of the
sum rule, allowing the deviation of $\chi^2$  by a factor 1.5 from its
minimal value, we find: $\Sigma=0.33\pm 0.20$, $\chi^{\prime}(0)=(2.4\pm
0.5)\cdot 10^{-3}~GeV^2$.

From the sum rule analogous (12) it is possible to find $g^8_A$ -- the
proton coupling constant  with the octet axial current, which enters the QCD
formula for $\Gamma_{p,n}$.

The $M^2$ -dependence of $g^8_A + C_8 M^2$ is presented in Fig.1a and
the best fit according to the fitting procedure (14) at $1.0 \leq M^2
\leq 1.3 ~ GeV^2$ gives
%21
\begin{equation}
g^8_A = 0.65 \pm 0.15, ~~~ C_8 = 0.10 ~GeV^{-2} ~~~\sqrt{\chi^2} = 1.2
\times 10^{-3}
\end{equation}
The obtained value of $g^8_A$
within the errors coincides with $g^8_A = 0.59 \pm 0.02$  found
from the data on baryon octet $\beta$-decays under assumption of strict
SU(3) flavour symmetry \cite{HSU}.

A similar sum rule with the account of dimension 9 operators can be
derived also for $g_A$ -- the nucleon axial $\beta$-decay coupling
constant. It  is an improvement of the sum rule found in \cite{KOG} and has 
the form 
%16 
\begin{equation} 
g_A + C_A M^2 = 1 + \frac{8}{9 
\tilde{\lambda}^2_N} e^{m^2/M^2} \Biggl [a^2 L^{4/9} + 2 \pi^2 m^2_1 
f^2_{\pi} M^2 - \frac{1}{4} a^2 \frac{m^2_0}{M^2} + \frac{5}{3} \pi \alpha_s 
f^2_{\pi} \frac{a^2}{M^2} \Biggr ] 
\end{equation} 
The $M^2$ dependence of 
$g_A - 1 + C_A M^2$ is plotted in Fig.1a, lower curve.  The best fit gives 
%17
\begin{equation}
g_A = 1.37 \pm 0.10, ~~~ C_A = -0.088~ GeV^{-2}, ~~~~ \sqrt{\chi^2} =
1.0 \times 10^{-3}
\end{equation}
in comparison with the world average $g_A
= 1.260 \pm 0.002$.

B.Ioffe is thankful to A.v.Humbolt Foundation for financial  support 
allowing him to participate at the Conference. 
The work was supported in part by
CRDF Grant RP2-132, RFFR Grant 97-02-16131 and
Swiss Grant 7SUPJ048716.

\begin{figure}
\epsfxsize=9cm
\epsfbox{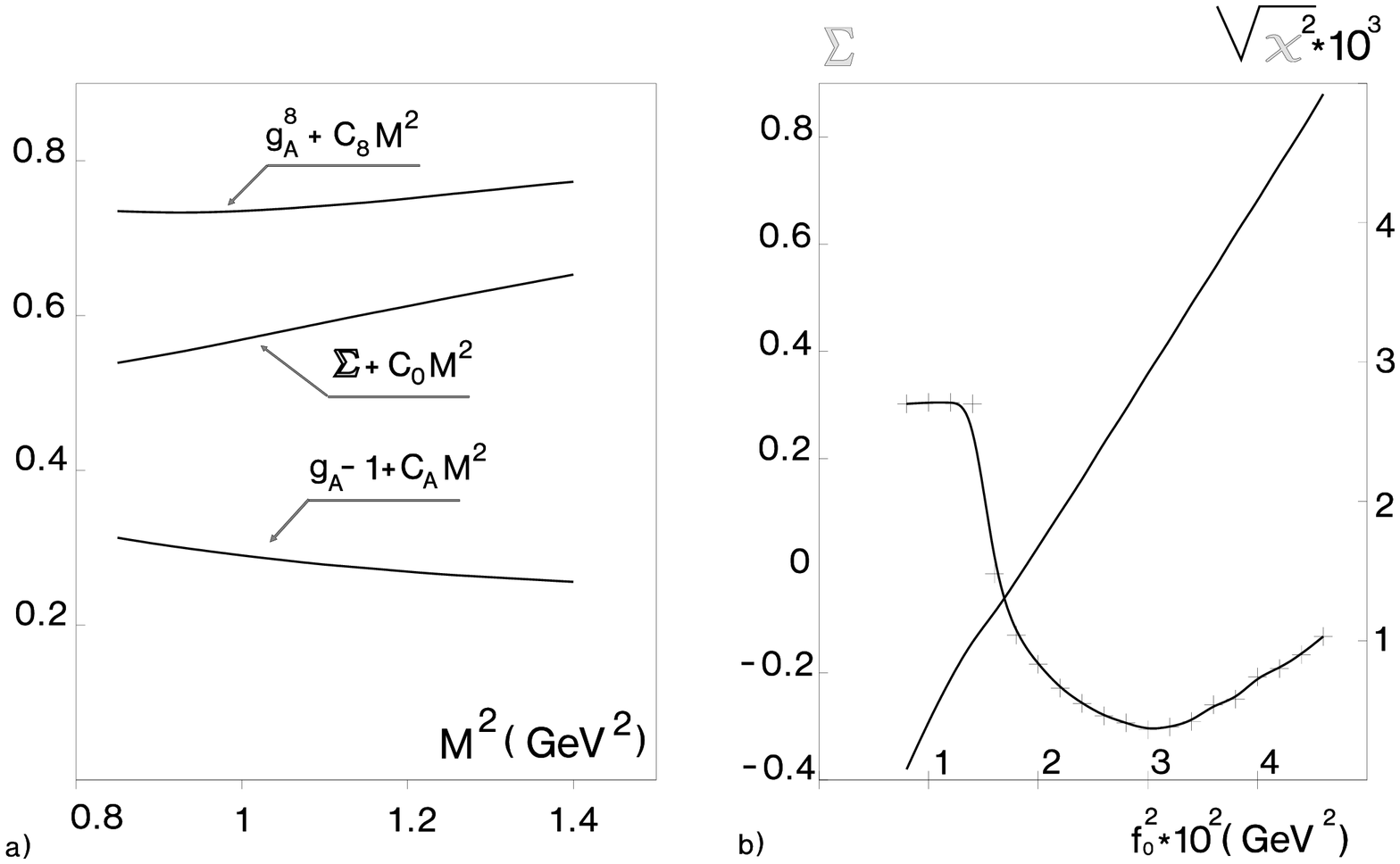}
\caption{a.) The $M^2$-dependence of $\Sigma + C_0 M^2$ at $f^2_0 =
3 \times 10^{-2}~GeV^2$, $g^8_A + C_8 M^2$, and $g_A - 1 + C_A M^2$.
b.)$\Sigma$  (solid line) and $\sqrt{\chi^2}$ (crossed line)
as a functions of $f^2_0$.}
\end{figure}

\end{document}